\documentclass[twocolumn,showpacs,preprintnumbers,amsmath,amssymb]{revtex4}
\usepackage{graphicx}% Include figure files
\usepackage{dcolumn}% Align table columns on decimal point
\usepackage{bm}% bold math
\begin{document}
%\baselinestretch{5}
%\tightenlines
\topmargin=-0.3cm
\title{Large Momenta Fluctuations Of Charm Quarks In The Quark-Gluon Plasma}
%\vspace{0.1in}

\author{S.~Terranova, D.M. Zhou and A. Bonasera 
 \footnote{Email:  terranova@lns.infn.it;  zhou@lns.infn.it;  bonasera@lns.infn.it}}
\affiliation{
  Laboratorio Nazionale del Sud, Istituto Nazionale Di Fisica Nucleare,
      Via S. Sofia 44, I-95123 Catania, Italy \\
}
%\maketitle

\begin{abstract}                
We show that large fluctuations of D  mesons kinetic  energy  (or momentum)
distributions might be a signature of a phase transition to the quark gluon plasma (QGP).
In particular, a jump in the variance of the momenta or kinetic energy, as a function of a control
parameter (temperature or Fermi energy at finite baryon densities) might be a signature for a 
first order phase transition to the QGP.   This behaviour is completely consistent with the order
parameter defined for a system of interacting quarks at zero temperature and finite baryon densities which shows a jump in correspondance to a first order phase transition to the QGP. 
The  $J/\Psi$ shows exactly the same behavior of the order parameter and of the variance of the D mesons.  We discuss  implications for
relativistic heavy ion collisions  within the framework of a transport model and possible hints for
experimental data.
\end{abstract}

\maketitle

\hspace{-\parindent}{\bf PACS : {\bf 12.39.Pn  24.85.1p } }\\[1ex]

%\section{Introduction}
The production of a new state of matter, the Quark-Gluon Plasma (QGP), can be obtained through ultra-relativistic heavy ion 
collision (RHIC) at CERN and at Brookhaven \cite{wong}.
QGP can be formed in the first stages of the collisions, and can be studied 
through the secondary particles produced.

Some features 
of the quark matter can be revealed by studying the properties of hadrons in a dense medium.
The particle $J/\Psi$ is a good candidate because the formation of the QGP might lead
to its suppression\cite{Jpsi}.  Here we want to show that in reality informations about the QGP
are carried by the charm quarks.  These quarks interact strongly with the surrounding matter and
as a result we have  a suppression of the  $J/\Psi$, but also large fluctuations of the charm quarks kinetic energies which could be revealed by the D mesons distributions or other charmed mesons or baryons.  To see this, we elaborate on 
 a semiclassical model which has an EOS resembling the well known 
properties of nuclear matter and its transition to the QGP at zero temperature and finite baryon 
densities already discussed in \cite{salv04}.
% we try not only to reproduce the known properties of nuclear matter starting from 
%the quark degrees of freedom \cite{refpov},also  how 
We simulate the nuclear matter which is composed of 
nucleons (which are by themselves  composite three-quark objects) and its 
dissolution into quark matter. In addition, for our system of colored quarks, we will 
show how the color screening
is related to the lifetime of the particle $J/\Psi$ in the medium.
In particular, we will see that the lifetime of the $J/\Psi$ as a function of density behaves
as an order parameter.   On exactly the same ground we show that the variances of the charm
quarks are large and they display a jump at the critical point for a first order phase transition.
Thus this quantity, similarly to the lifetime of the $J/\Psi$, behaves exactly as an order parameter
and could give important informations about not only the transition to the QGP but also to the
order of the phase transition, i.e. first, second order or simply crossover to the QGP.  Of course since D particles 
are the lighest charmed meson they are most easily produced in heavy ion collisions thus they are the best probes
for the phase transition. 

 An important ingredient of our approach is a constraint to satisfy the
Pauli principle.  The approach, dubbed Constrained Molecular Dynamics (CoMD) has been 
successfully applied to relativistic and non-relativistic \cite{bon2000,papa} heavy ion collisions
 and plasma physics as well \cite{fus03}.

%\section{Numerical Method}
We use  molecular dynamics with a constraint for a Fermi
system of quarks with colors.
The color degrees of freedom of quarks are taken into account through the Gell-Mann 
matrices and their dynamics are solved classically, in  phase space, following
the evolution of the distribution function.  

In our work, the quarks interact through the Richardson's potential $V({\bf r}_i,{\bf r}_j)$: 
 \begin{eqnarray}
 V({\bf r}_{i,j})=3\sum_{a=1}^{8}\frac{\lambda_i^a}{2}\frac{\lambda_j^a}{2}\left[
\frac{8\pi}{33-2n_f}\Lambda(\Lambda r_{ij}-\frac{f(\Lambda r_{ij})}
{\Lambda r_{ij}})\right],
 \end{eqnarray}
and\cite{rich}
 \begin{eqnarray}
f(t)=1-4 \int{\frac{dq}{q}\frac{e^{-qt}}{[{\rm ln}(q^2-1)]^2+\pi^2}} .
 \end{eqnarray}
$\lambda^a$ are the Gell-Mann matrices. We fix the number of flavors 
$n_f=2$ and
the parameter $\Lambda=0.25$ GeV,($\hbar,c=1$) unless otherwise stated. 
Here we assume the potential to be dependent on the relative coordinates only.
The first term is the linear term, responsible for the confinement, the second is 
the Coulomb term \cite{rich}.
 
We solve the classical  Hamilton's equations:
\begin{equation}
\label{efer} 
\frac{d{\bf r}_i}{dt}=\frac{{\bf p}_i}{E_i} ,
\end{equation}
\begin{equation}
\label{efer}
\frac{d{\bf p}_i}{dt}=-\overrightarrow{\nabla}_{{\bf r}_{i}} U({\bf r}).
\end{equation}

Initially we distribute randomly the quarks in a box of side $L$ in coordinate space and
in a sphere of radius $p_f$ in momentum space.   
$p_f$ is the Fermi momentum
estimated in a simple Fermi gas model by imposing that a cell in
phase space of size $h=2 \pi$ can accommodate at most $g_q$ identical quarks
of different spins, flavors and colors.
$g_q=n_f\times n_c\times n_s$ is the degeneracy number, $n_c$ is the number
of colors (three different colors are used: red, green and blue ) hence $n_c=3$;
$n_s=2$ is the number of spins\cite{wong}. 
 
A simple estimate gives the following relation between the density of quarks with colors,
$\rho_{qc}$, and the Fermi momentum:
\begin{eqnarray}
\rho_{qc}=\frac{3n_s}{6\pi^2}p_f^3 
 \end{eqnarray}

We generate many events and take the average over all events in each cell on the phase space.
For each particle we calculate the occupation average, i.e. the probability that a cell in the
phase space is occupied.   To describe the Fermionic nature of the system we impose
 that average occupation for each particle is less or equal to 1 $(\bar{f_i}\leq 1)$.

At each time step we control the value of the average distribution function and consequently we 
change the momenta of particles by multiplying them by a quantity $\xi$:
$P_i=P_i\times \xi$.
$\xi$ is greater or less than 1 if $\bar{f_i}$ is greater or less than 1 respectively; 
which is the $constraint$ \cite{papa}.  Details of the model are discussed in \cite{salv04}.
Here we illustrate the case with  $m_u =5$ MeV, $m_d =10$ MeV and cut-off to the linear term 
of $2fm$ \cite{salv04} which has been 
introduced to avoid numerical uncertainties. Such a system displays a clear first order phase transition at high baryon densities.  The results are completely analogous for the other parameter sets discussed in \cite{salv04}. 

%%%%%%%%%%%%%%%%%%%%%%%%%%%%%%%%%%%%%%%%%%%%%%%%%%%%%%%%%%%%%%%%%%
%%% Fig 1 %%%% temporal evolution
%%%%%%%%%%%%%%%%%%%%%%%%%%%%%%%%%%%%%%%%%%%%%%%%%%%%%%%%%%%%%%%%%%
\begin{figure}[htbp]
\centering
 \includegraphics[width=8cm,clip]{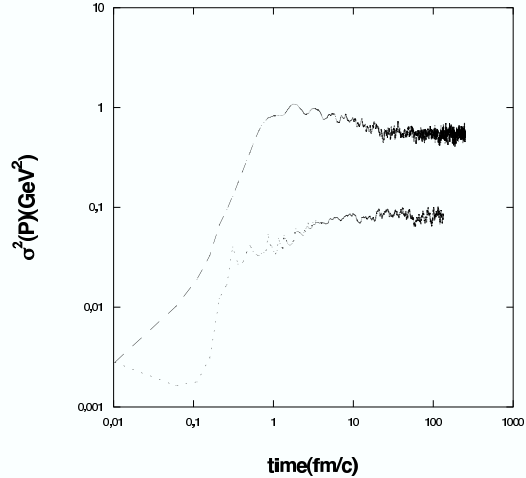}% Here is how to import EPS art
\vspace{0.1in}
\caption{Time evolution of momentum variances of charm quarks
 at two densities, above and below the QGP phase transition.} 
\label{time}
\end{figure}
%%%%%%%%%%%%%%%%%%%%%%%%%%%%%%%%%%%%%%%%%%%%%%%%%%%%%%%%%%%%%%%%%%%

After our system of u and d quarks has evolved to its equilibrium configuration, we insert one
 $J/\Psi$, i.e.  c $\bar c$ quarks and let them evolve.  The Pauli principle is responsible for the kinetic energy of the light quarks.  The Fermi motion increases for increasing densities.  On the
other hand the embedded charm quarks do not see the Pauli principle because they are different
Fermions.  However, there is a strong interaction among all the quarks given by the Richardson potential.   Because of such interaction, the charm quarks  start to exchange energy with the surrounding medium and finally get in equilibrium with
the other quarks.  Thus c quarks are a perfect probe of the system since they can wander anywhere in 
the available phase space of the system.  In particular we can calculate
the variance for instance of the charm quarks.  For a non interacting Fermi gas, the variance in 
momentum space is given by
\begin{eqnarray}
\sigma^2=<p^2>-<p>^2=\frac{3}{80}p^2_F 
 \end{eqnarray}

 similar relations can be obtained  for kinetic energy fluctuations.  Of course, because of the interaction, the fluctuations
are larger than our estimate and might spectacularly increase near a phase transition.  

In fig.(1),
we plot the variance in momentum space vs time for two densities above and below the
critical point for a first order phase transition.  One immediately sees that variances are much 
 larger 
than our estimate for a Fermi gas (here respectively $\sigma^2=0.0055$ and $0.0328 GeV^2$ 
for the cases diplayed in fig.1)
 given above because of the strong interaction.  
Initially the variance of the c quarks
are very small, but after a transient time, fluctuations are transferred from the light to the heavy quarks up to a stationary value.  As expected the variances are larger above the phase transition,
i.e. at high density.   
%%%%%%%%%%%%%%%%%%%%%%%%%%%%%%%%%%%%%%%%%%%%%%%%%%%%%%%%%%%%%%%%%%
%%% Fig 2 %%%%  Lifetime of J/psi
%%%%%%%%%%%%%%%%%%%%%%%%%%%%%%%%%%%%%%%%%%%%%%%%%%%%%%%%%%%%%%%%%%
\begin{figure}[ht]
\centering
  \includegraphics[width=8cm,clip]{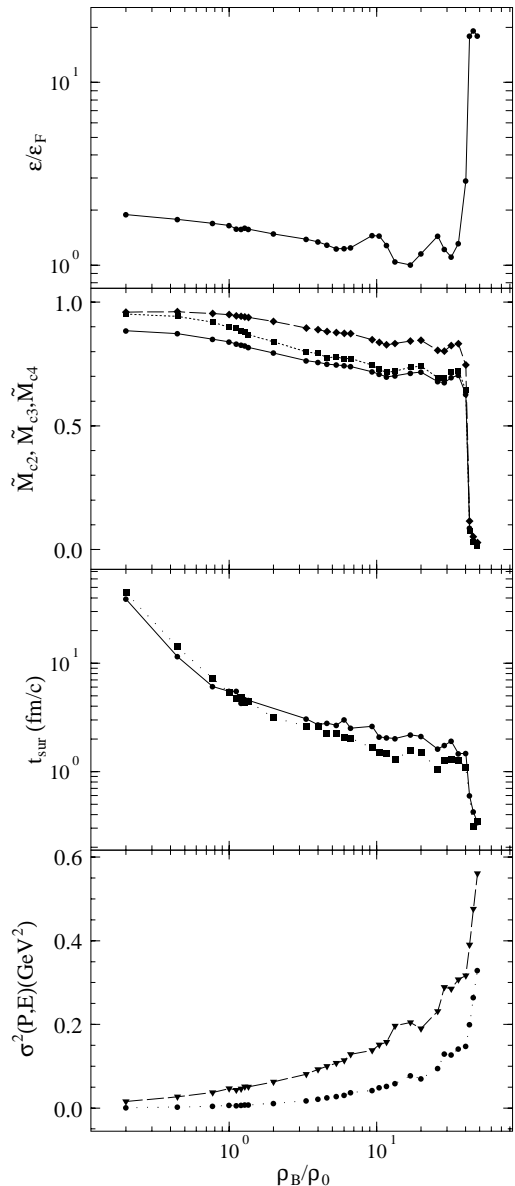}% Here is how to import EPS art
\vspace{0.2in}
\caption{Energy density (top panel), normalized order parameters
(2nd panel), time survival of $J/\psi$ (3rd panel) and variances in momentum and kinetic energy 
of c quarks (bottom panel) versus density
divided by the normal density $\rho_0 $, for $m_u =5$ MeV, $m_d =10$ MeV 
and cut-off$=2fm$.} 
\label{life}
\end{figure}

This is clearly demonstrated in figure (2) where the energy density, the order parameters, the $J/  \Psi $ lifetime \cite{salv04} and the variances are plotted versus density.  All quantities
jump at the critical point signaling a first order phase transition.  In order to show that these 
properties are independent on details of the forces and that the charm quarks are real good 
probes of the phase transition, we have arbitrarily increased of a factor 2 the interaction strenght 
between the c$\bar c$ quarks alone.  This results in a change of the $J/\Psi$ lifetime (squares in
fig.2) but the jump of its lifetime at the critical point remains.

%%%%%%%%%%%%%%%%%%%%%%%%%%%%%%%%%%%%%%%%%%%%%%%%%%%%%%%%%%%%%%%%%%
%%% Fig 3 %%%% fluctuations
%%%%%%%%%%%%%%%%%%%%%%%%%%%%%%%%%%%%%%%%%%%%%%%%%%%%%%%%%%%%%%%%%%
\begin{figure}[ht]
\centering
  \includegraphics[width=8cm,clip]{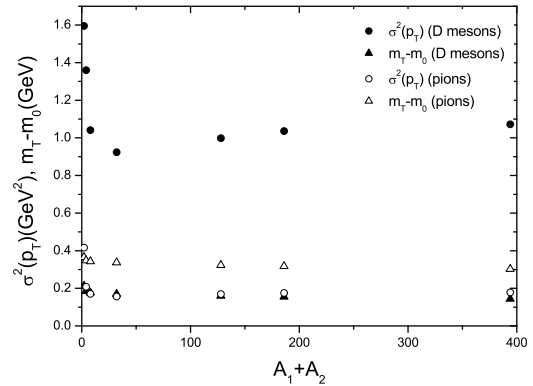}% Here is how to import EPS art
\vspace{0.1in}
\caption{Variances (circles) and transverse minus rest mass (triangles) vs mass number of the colliding nuclei.
Full symbols refer to D-meson production while open symbols refer to pions.} 
\label{vari}
\end{figure}
%%%%%%%%%%%%%%%%%%%%%%%%%%%%%%%%%%%%%%%%%%%%%%%%%%%%%%%%%%%%%%%%%%%

In order to see what could happen in  realistic heavy ion collisions we have performed some
calculations in a transport model at $\sqrt{s}=200 GeV$.  We have used the JPCIAE code which
includes Pythia as a generator of elementary collisions \cite{sa} and performed calculations
for symmetric nucleus-nucleus collisions starting from pp to AuAu.  Variances in the transverse
 momenta of produced open charms are calculated for central collision (impact parameter b=0fm).
 In figure (3) we plot the variances versus sum of the mass numbers of the colliding nuclei.  The
variances are calculated averaging over the events which contain at least one D meson produced (circles).
In the same graph we have plotted the transverse mass minus the rest mass of the particle(triangles),
 which gives an indication
of the degree of equilibration reached.  The full symbols refer to D-mesons while the open symbols refer to pions.
 As expected such variances are roughly a constant i.e. independent of the mass number for colliding ions heavier
than oxigen.  Infact, this is an expected result since the model contains in principle no phase transition
 to the QGP.   Thus normalizing the variance say in Au+Au respect p+p collisions at the
same beam energy should give about 1.  An anomalous increase
 from such a value should indicate
that much more physics than contained in our transport code
is indeed at play.  Furthermore, some anomaly in the variances with increasing mass number
of the colliding nuclei should give a clear indication to the occurrence of the phase transition 
and, possibly, of its order.  This is so because we expect no QGP for small systems while a 
transition should occur for relatively large nuclei.  How large those nuclei should be at RHIC
 beam energy could be obtained from the analysis we propose in figure(3).   In the same figure we have plotted
the average transverse kinetic energy for open charms and pions.  Such a quantity should give an indication of the
degree of equilibration of the system.  As it can be easily seen the latter quantity has a similar behaviour of the
variance which is what we expect if the system is equilibrated.  Indeed one can see that there is a proportionality between
the two quantities especially for colliding nuclei heavier than oxigen, furthermore the average kinetic energies for different
particles are very close which is another indication of thermal equilibrium.  Looking at the pions results only, it seems that
equilibrations is reached already for small systems such as d+d.  This is so because many pions are produced in an event and also,
pions are produced in different steps of the collision.  This results in smaller variances than the one obtained for D-mesons already
in our kinetic approach.  Thus pions could give an indications on the temperature reached in the reaction, while a phase transition
could be signaled by the possible $J/\Psi$ suppression or the large fluctuations of the D-meson which might be
 a stronger signal since they are more abundantly produced in the collisions.  In this context, large fluctuations are
understood compared to the pp case.

%\section{Summary}
In conclusion, in this work we have discussed a semiclassical molecular dynamics approach to 
infinite matter at finite baryon densities and zero temperature starting from a phenomenological 
potential that describes the interaction between quarks with color.  Pauli blocking, necessary for Fermions at zero temperature, is enforced through a constraint to the average one body 
occupation function.
We have studied the case of a set of parameters which displays a first order phase transition at high baryon densities.   We have shown that,
similarly to the order parameter and the $J/\Psi$ lifetime, the variances in momentum space of charm quarks 'jump' at the critical point
of the phase transition.   We have proposed an experimental search at RHIC inspired by the CoMD results.  In particular we have 
simulated heavy ion collisions in a transport model (which does not include a phase transition),
 at fixed beam energy and changing the mass numbers of the colliding system.  We have shown that
already for small colliding nuclei such as oxigen the system could reach thermal equilibrium. An anomalous behavior of the variances of the
open charms transverse momenta as a function of the mass number of the colliding system should give a signal for the transition to the
QGP.  In particular variances for pp collisions should be much smaller than those obtained in Au+Au collisions at the same beam energy if 
there is a phase transition.  If not, we expect variances to be in agreement to our estimate in fig.(3).  If the large fluctuations are found in 
open charms production, then the phase transition could be further studied by changing the beam energy and/or the 
impact parameter selection for a given system and energy, 
since we expect a phase transition for central and not for the most peripheral
collisions.
\vskip 0.7cm
\begin{center}
{\bf Acknowledgements}
\end{center}
We thank Prof. R.Rapp for stimulating discussions and prof. B.H.Sa for making his transport code available to us.

\end{document}